\documentstyle[prl,aps,epsf,floats]{revtex}

\begin{document}

\twocolumn[\hsize\textwidth\columnwidth\hsize\csname @twocolumnfalse\endcsname
\author{A.M. Bratkovsky$^{1}$ and A.P. Levanyuk$^{1,2}$}
\address{$^{1}$ Hewlett-Packard Laboratories, 1501 Page Mill Road, Palo
Alto, California 94304\\
$^{2}$Departamento de F\'{i}sica de la Materia Condensada, C-III,
Universidad Aut\'{o}noma de Madrid, 28049 Madrid, Spain
}
\title{
Phase transitions and ferroelectricity in very thin films: single- 
versus multidomain state 
}
\date{January 14, 2006}
\maketitle
\begin{abstract}
We discuss ferroelectric phase transitions into single- and
multidomain states in very thin films using continuous theory. It is shown that
in nearly cubic ferroelectrics the domain state may survive down 
to atomic film thicknesses, unlike the single domain state, which is almost always 
unstable or metastable. This conclusion 
is valid almost irrespective of the nature of electrodes
(metallic or semiconducting) and whether or not the screening carriers
may be present in the ferroelectric itself.
\end{abstract}

\vskip2pc] \narrowtext

With a thrust towards developing nanoelectronics components, like
ferroelectric (FE) memories\cite{scott00} the question of phase
transformation and the very existence of ferroelectricity (problem of
``critical thickness'', first raised decades ago \cite
{ivanchik61,guroFmet70,batra72,wurfel74}) becomes ever more important. It is
a focus of experimental and theoretical research, see e.g. \cite
{noh05,ghosez03}. Here we will present the thermodynamic results for phase
behavior of ferroelectric thin films with various electrodes and consider
both homogeneous and inhomogeneous states of the film as a function of
temperature and thickness. Many of prior works have given some results
almost exclusively for homogeneous (monodomain) ferroelectric film states 
\cite{ivanchik61,guroFmet70,batra72,wurfel74,wat} with estimates for a
``critical thickness'' ranging from $\sim 1\mu $m\cite{ivanchik61} to a few
nm \cite{wat}. We shall call critical the thickness where either homogeneous
or inhomogeneous spontaneous polarization becomes unfavorable. Obviously, it
is non-universal, and depends on many details like the ferroelectric
material, lattice mismatch, type of electrodes, etc. For instance, we show
below that the homogeneous state studied in Ref.~\cite{ghosez03} may exist
only when very special conditions are met, in most real situations the FE
films are likely metastable or even unstable with regards to breaking into
domains. Importantly, ferroelectric domains in nearly unstrained cubic FE
films may exist down to practically one unit cell.

Consider a slab of uniaxial ferroelectric occupying the region $-l/2<z<l/2$
between two metallic (or degenerate semiconductor)\ electrodes. The polar
axis is perpendicular to the film plane (parallel to the $z-$axis), and the
bias voltage is set as $-(+)U/2$ at the right (left) electrode$.$ The
electric field would penetrate into electrode over very short Thomas-Fermi
screening length\cite{Mead61}, and this screening layer manifest itself as
some additional ``interface capacitance''\cite{tilley96}. To characterize
the screening, we introduce the {\em band} {\em bending }potential $\phi $
as $\varphi =\phi \mp U/2$. Since the screening is assumed to be very
efficient, the potential drop in the electrode is small compared to the
Fermi energy $\mu $, $|q\phi |/\mu \ll 1,$ and the Poisson equation
linearizes to $\phi ^{\prime \prime }=\kappa ^{2}\phi ,$ where $\kappa =%
\sqrt{6\pi nq^{2}/\epsilon _{s}\mu }$ is the inverse screening length, $%
\kappa ^{-1}=0.5-0.7$\AA\ in metals\cite{Mead61}, comparable to the atomic
distance $d_{at},$ $n$ the net carrier density, $\epsilon _{s}$ the
dielectric constant, $q$ the elementary charge. According to this equation,
the band bending is antisymmetric, $\phi (-z)=-\phi (z),$ $\phi =\phi
_{1}e^{-\kappa (z-l/2)}$ at $z>l/2$, with 
\begin{equation}
\phi _{1}=\left( 2\pi Pl+U/2\right) /(1+\epsilon _{s}\kappa l/2).
\label{eq:fi1}
\end{equation}
This is the local electrodynamic {\em boundary condition}. In fact, the
surface has properties different from the bulk and generally{\bf \ }produces
an effective ``field''\ $w$ coupled to the normal component of polarization,
as was pointed out recently\cite{BL05}. This, in principle, requires to add
to (\ref{eq:fi1}) the {\em additional boundary conditions }for polarization 
\cite{degennes69,binder,tilley96,BL05}. The solution (\ref{eq:fi1})\ applies
when $|q\phi _{1}|\ll \mu ,$ that translates for metal, where $\kappa l\sim
l/d_{at}\gg 1,$ into $4\pi q|P|/\kappa \sim 4\pi q|P|d_{at}\ll \mu ,$ which
is always satisfied. The condition would be violated in moderately/lightly
doped semiconductors.

The equation of state of a ferroelectric 
\begin{equation}
AP+BP^{3}=E,  \label{eq:meq}
\end{equation}
where the homogeneous\ field in the monodomain\ ferroelectric would be $%
E=\left( \epsilon _{s}\kappa U-8\pi P\right) /\left( \epsilon _{s}\kappa
l+2\right) ,$ $A=(T-T_{c})/T_{0}$, $T_{c}$ is the critical temperature, $%
T_{0}$ the characteristic temperature, $T_{0}\sim T_{at}$ in displacive
ferroelectrics, where $T_{at}\sim 10^{4}-10^{5}$K is the characteristic
``atomic'' temperature. The field $E=E_{0}+E_{d},$ where $E_{0}=U/l$ is the
external and 
\begin{equation}
E_{d}=-PL_{0}/l  \label{depfield}
\end{equation}
is the depolarizing field, $L_{0}=8\pi /\epsilon _{s}\kappa $ the
characteristic length scale in electrodes, and we have used the fact that in
a metal the screening length is small, $\kappa l\gg 1.$ The homogeneous
ferroelectric state in short-circuited electrodes, $U=0,$ requires an
intersection of the $P\left( E\right) $\ curves given by Eqs.~(\ref{eq:meq})
and (\ref{depfield}) at $P\neq 0$ (Fig.~1).{\bf \ }This necessary, but not
sufficient, condition is realized at 
\begin{equation}
A<A_{h}\equiv -L_{0}/l,  \label{eq:Ah}
\end{equation}
illustrated in Fig.~1a, where curves 1-3 are plotted for $T_{1}>T_{2}>T_{3}.$
The nontrivial solutions exist only for $T_{2,3}$. If these solutions were
stable with respect to formation of inhomogeneous states, the corresponding
critical temperature would be suppressed by $\Delta T_{c}=-T_{0}L_{0}/l\sim
8\pi T_{0}d_{at}/\epsilon _{s}l,$ compared to the bulk $T_{c}$\cite
{batra72,tilley96}. Hence, the {\em homogeneous} ferroelectricity would
become impossible in films with a thickness below 
\begin{equation}
l_{hc}=8\pi /\epsilon _{s}\kappa |A|_{\max }=L_{0}T_{0}/T_{c},
\end{equation}
where $|A|_{\max }=T_{c}/T_{0}.$ In displacive systems $|A|_{\max }\sim
10^{-2}-10^{-3}$, so for $\epsilon _{s}\sim 1-10$ and $\kappa _{m}\sim
d_{at}^{-1}$ we find $l_{hc}\sim \left( 10^{2}-10^{3}\right) d_{at}\sim
50-500$\AA $.$ One should bear in mind that the ferroelectricity was
observed in thinner films \cite{noh05} but it was not possible to check if
the film was mono- or polydomain, the latter being much more likely
scenario. Obviously, when the external field is larger than the depolarizing
field (\ref{depfield}), $E_{0}\geq E_{d},$ the film will be in monodomain
state. This indicates the boundary between poly- and monodomain states and
correctly describes the tilt of the hysteresis loops with decreasing film
thickness observed in Ref.\cite{noh05}. 

\begin{figure}[t]

\epsfxsize=3in 
\epsffile{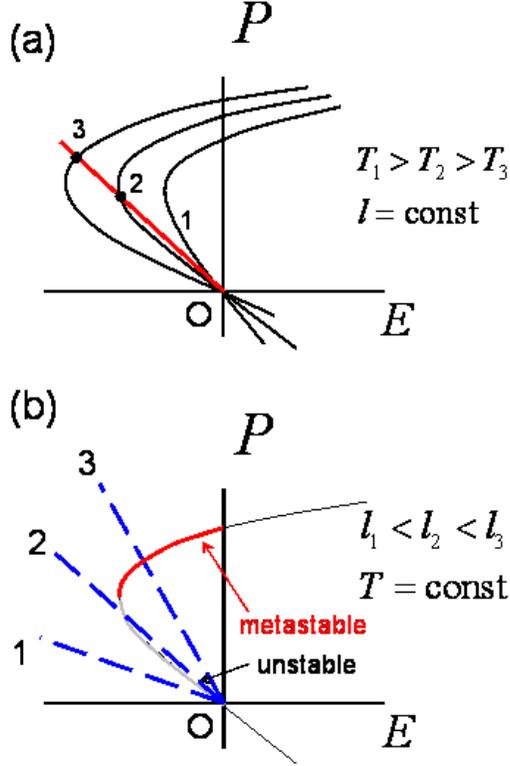 }

\caption{ Solutions of the equation of state for FE film (\ref{eq:meq}) with
the depletion field $E_d$ (\ref{depfield}): (a) at constant thickness of the
film $l$ and decreasing temperature; (b) at constant temperature $T$ and
varying thickness $l$. Nontrivial solutions, when they exist, are either
unstable (2) or metastable (3) with respect to domains. }
\label{fig:1}
\end{figure}

The necessity to study the sufficient conditions is suggested by Fig.~1,
since the homogeneous ferroelectric solution belongs in the regions which
correspond to either unstable or metastable states in the bulk samples. The
problem of stability loss with respect to small fluctuations is a nontrivial
yet tractable (linear) problem \cite{Chensky82,BLinh}. The check on
metastability requires the calculation of free energies, which are much more
difficult to find, so only limiting cases will be discussed below. The
question of stability with respect to small fluctuations splits into two
parts: stability with respect to (i) homogeneous and (ii) inhomogeneous
fluctuations. Any solution of Eqs.~(\ref{eq:meq}) and (\ref{depfield}) is
stable with regards to homogeneous fluctuations. One can prove this by
solving a relaxation dynamics for polarization constructed by generalizing
the equation of state (\ref{eq:meq}): 
\begin{equation}
\gamma \frac{\partial P}{\partial t}=-\tilde{A}P-BP^{3},  \label{eq:dP}
\end{equation}
where the relaxation parameter $\gamma >0$. Linearizing this equation about
the point $P=P_{0},$ where 
\begin{equation}
P_{0}=\sqrt{-\tilde{A}/B},\quad \tilde{A}=A+L_{0}/l,  \label{eq:P01}
\end{equation}
one obtains for homogeneous fluctuations $\delta ^{h}P=P-P_{0}$ 
\begin{equation}
\gamma \frac{\partial \delta ^{h}P}{\partial t}=-\left( \tilde{A}%
+3BP_{0}^{2}\right) \delta ^{h}P,  \label{eq:dPdt}
\end{equation}
so that the perturbation does indeed {\em decay} with time as $\delta
^{h}P\propto e^{\alpha t}$ with the decrement $\alpha =2\tilde{A}<0$.

Now let us discuss stability with respect to {\em inhomogeneous}
fluctuations of polarization. The present situation with the field
penetrating into electrode over the Thomas-Fermi length is analogous to a
ferroelectric film separated from the electrodes by nonferroelectric
``dead'' layers\cite{BL00,BLdlay01,Chensky82} that promote breaking the film
into domains. To obtain the conditions for domain instability in the present
case, one looks for a nontrivial solution of the equation of state with the
gradient term included\cite{BLinh} 
\begin{equation}
AP+BP^{3}-g\nabla _{\perp }^{2}P=E,  \label{eq:Eqstateinh}
\end{equation}
where $\nabla _{\perp }^{2}=\partial _{x}^{2}+\partial _{y}^{2}$ is
``in-plane'' Laplacian, together with Maxwell equations. The solution is
sought in a form of a ``polarization wave'' $P=P_{0}+\delta
P_{k}(z)e^{ikx},\qquad \varphi =\varphi _{k}(z)e^{ikx},$ where $\delta
P_{k}(z)$ is the small perturbation over the homogeneous polarization $P_{0} 
\cite{Chensky82,BLinh}$. In the case of metallic screening we obtain the
following condition for an existence of the nontrivial solution with certain 
$k:$ 
\begin{equation}
\chi \tan \frac{1}{2}\chi kl=\frac{\epsilon _{\perp }k}{\epsilon _{s}\sqrt{%
k^{2}+\kappa ^{2}}},  \label{eq:chi}
\end{equation}
where $\chi ^{2}=-\epsilon _{\perp }(\hat{A}+gk^{2})/4\pi >0,$ $\hat{A}%
=A+3BP_{0}^{2}$. The case of interest to us is $\kappa \gg k,$ which is easy
to meet in the present case of metallic electrodes. We assume (and check
validity later)\ that $\epsilon _{\perp }k/\epsilon _{s}\kappa \chi \gtrsim
1.$ Then, the equation simplifies to $\chi kl=\pi ,$ the same as in FE\ film
without electrodes. Substituting there the above expression for $\chi (k,%
\hat{A}),$ we find the maximal value of $\hat{A}_{c}=-2gk_{c}^{2}$ at $%
k=k_{c}$ where this equality is first met, defining the temperature where
the instability sets in. The ``polarization wave'' forms at $\hat{A}<0$ such
that 
\begin{eqnarray}
-\hat{A}_{c} &=&2gk_{c}^{2}=-A_{d}=\lambda /l,  \label{eq:Ad} \\
k_{c} &=&\left( \frac{4\pi ^{3}}{\epsilon _{\perp }gl^{2}}\right) ^{1/4}\sim 
\frac{1}{\epsilon _{\perp }^{1/4}\sqrt{d_{at}l}},  \label{eq:kc}
\end{eqnarray}
where $g$ is the coefficient before the gradient term in the equation of
state (\ref{eq:Eqstateinh}), $\lambda =4\pi ^{3/2}g^{1/2}/\epsilon _{\perp
}^{1/2}\sim d_{at}/\epsilon _{\perp }^{1/2}$ is the characteristic thickness
for the domains formation. Now, checking the assumption that we used to
solve the Eq.~(\ref{eq:chi}), we see that it boils down to $\sqrt{4\pi
\epsilon _{\perp }}/\left( \epsilon _{s}\sqrt{\kappa ^{2}g}\right) \sim 
\sqrt{4\pi \epsilon _{\perp }}/\epsilon _{s}\gtrsim 1,$ which is met in
(e.g. perovskite)\ films with $\epsilon _{\perp }\sim 100-1000,$ and the
typical $\epsilon _{s}\sim 1-10.$

These results remain basically unchanged if one were to account for
realistic boundary conditions and the fact that the interface creates the
effective ``field'' $w$ coupled to the normal polarization component\cite
{BL05}. The surface field results in a ``frozen''\ polarization $P_{0}(z),$
but since it extends over a short length on the order of a lattice spacing,
it makes the film response ``hard''\ just near the boundary. The film
remains ``soft'' in the bulk and there the stability loss proceeds by the
above scenario via appearance of the polarization wave. The effect of the
surface field is basically that the surface layer of atomic thickness would
be excluded from the process of domain formation.

Importantly, in nearly unstrained cubic ferroelectric films the multidomain
states are much more likely than in standard uniaxial ferroelectrics.
Indeed, upon cooling from the paraelectric phase of a uniaxial ferroelectric
film a{\em \ }domain instability sets in a form of\ a ``polarization wave''
when $A=A_{d}$ (\ref{eq:Ad}) with the critical wave vector $k=k_{c}$ (\ref
{eq:kc})\ found from 
\begin{equation}
|A_{d}|_{\text{uniaxial}}=4\pi ^{3/2}g^{1/2}/\epsilon _{\perp }^{1/2}l,
\label{eq:Acuni}
\end{equation}
Eq.~(\ref{eq:Ad}), where $\epsilon _{\perp }=1+4\pi /A_{\perp }$ is the
dielectric constant in the direction perpendicular to ferroelectric axis in
the plane of the film. This is the case of interest to us, since $\epsilon
_{\perp }\sim 10^{3}$ in BaTiO$_{3}$ at 2\% lattice misfit\cite
{pertsev98,ghosez03}. In standard situations Eq.(\ref{eq:Acuni})\ applies,
but if cubic perovskite films are grown on a substrate with small to
negligible lattice mismatch, the period of a domain structure in a cubic
ferroelectrics would increase. At the limit of applicability of this formula
one can estimate $A_{\perp }\approx A_{\parallel }\sim gk_{c}^{2}$ and we
will call them ``near-cubic''. Substituting this into (\ref{eq:Acuni}), we
obtain $k_{c}\sim \pi /(2^{1/2}l)\sim 1/l,$ so that the domain width $a\sim
\pi k_{c}^{-1}$ tends to become comparable to the thickness of the film. In
this case, according to (\ref{eq:Acuni}), the transition occurs very near
the bulk critical temperature at 
\begin{equation}
|A_{d}|_{\text{near cubic}}\sim g/l^{2}\sim (d_{at}/l)^{2}.  \label{eq:Anc}
\end{equation}
The minimal thickness, where the near cubic thin films can still transform
into a polydomain ferroelectric state, is then estimated for e.g. BaTiO$_{3}$
as $l_{cd}\sim d_{at}\left( \max |A_{d}|_{\text{near cubic}}\right)
^{-1/2}\sim 6d_{at}\sim 10$\AA\ at low temperature. Therefore, the
ferroelectricity in near cubic ferroelectric films can exist in a polydomain
form down to ``atomic'' thicknesses (just about one unit cell thick), where
the present continuous theory is at the border of validity but still able to
produce semiquantitative results. One can estimate from (\ref{eq:Acuni})\
that this regime corresponds to $\epsilon _{\perp }\approx 4l^{2}/g\sim
(l/d_{at})^{2}$. A more careful consideration suggests that the transition
in a cubic ferroelectric film with a {\em smaller} lattice misfit with the
substrate\ than the above borderline value proceeds into a {\em monodomain
state} with homogeneous {\em in-plane} polarization, e.g. $P_{x}$. This
threshold misfit is very small indeed, so in most cases the film remains
uniaxial and splits into domains according to the above scenario.

One can identify two possible cases from Eqs.~(\ref{eq:Ah}) and (\ref{eq:Ad}%
), while considering possible phase transitions upon lowering {\em %
temperature}: (i) $A_{h}>A_{d},$ when $\lambda >L_{0},$\ the phase
transition is into a homogeneous state, and (ii) $A_{h}<A_{d}$ when $\lambda
<L_{0},$\ the phase transition proceeds into a multidomain state. The case
(i)\ does {\em not} mean that the state remains homogeneous. It is stable
with respect to inhomogeneous fluctuations, but when the temperature
approaches $A\approx A_{d},$ domains would start to form. Indeed, at this
point the domain wall thickness $W\sim \sqrt{-g/A_{d}}$ becomes smaller to
the domain width $a,$ $W\lesssim a,$ and the existence of usual domains
becomes possible. The situation with metallic electrode is analogous to a
system with dead layers, which is metastable far from the critical point at $%
\left| A\right| \ll \left| A_{d}\right| $\ with respect to domain structure
at any thickness of the dead layer\cite{BL00}, and the domains are likely to
form already at $A\approx A_{d}.$ Consider now the case (ii) $\lambda
<L_{0}, $ where paraphase becomes unstable with respect to domains close to
the phase transition point. Interestingly, it was found in Ref.~\cite
{Chensky82} that the homogeneous state may be stable with respect to small
inhomogeneous fluctuations not only in the paraphase but also in the
ferrophase at some $T<T_{s}.$ This result is formal, however, since the
system is likely metastable in this region and domains will grow. The
temperature $T_{s}$ is defined by the condition $A+3BP_{0}^{2}=A_{d}=-%
\lambda /l.$ For the case $g=\lambda =0$ this corresponds to a point where a
depolarizing field equals thermodynamic coercive field, separating bulk
homogeneous metastable from unstable states. At $T<T_{s}$ and negligible
energy of the domain walls ($g=0)$ the depolarizing field would split the FE
film into domains. Therefore, the homogeneous state below $T_{s}$ is
actually {\em metastable}, and this is also easy to prove by comparing its
free energy with the domain state. The case with $g\neq 0$ is more involved,
but should be qualitatively similar, so that the low-temperature homogeneous
state predicted in Ref.~ \cite{Chensky82} is actually unobservable in the
thermodynamic sense.

It has been found in first-principles modeling of BaTiO$_{3}$ ferroelectric
film with SrRuO$_{3}$ metallic electrodes with in-plane lattice parameter
corresponding to a SrTiO$_{3}$ substrate (i.e. mimicking a capacitor
structure grown on top of SrTiO$_{3}$ with $\sim 2\%$ compressive strain)
that the state with a {\em homogeneous} polarization in $c-$direction
remains stable down to $\sim 24$\AA\ at zero temperature \cite{ghosez03} ($%
\sim $10\AA\ in unstrained PbTiO$_{3}$\cite{rappe05})$.$ Further, it was
claimed that the {\em depolarizing field} was solely responsible for
vanishing of the ferroelectricity in thinner films, the chemistry of the
interface (within the present nomenclature: the additional boundary
conditions, ABC) was deemed unimportant (however, see \cite{rappe05}).

With regards to such phase transitions {\em with thickness} of the film,
there are again two possibilities for samples with the thickness $%
l=l_{h}=L_{0}/|A|$ where $P_{0}=0$: (i) paraelectric state with zero
polarization is stable if $A=A_{h}>A_{d}$ (i.e. $L_{0}<\lambda ),$ and (ii)\
it is unstable and domains form, $L_{0}>\lambda .$ The condition (i) is met
when 
\begin{equation}
\sqrt{4\pi \epsilon _{\perp }}/\left( \epsilon _{s}\sqrt{\kappa ^{2}g}%
\right) <1.  \label{eq:C1}
\end{equation}
For very moderate values for perovskites $\epsilon _{\perp }\sim 100-1000,$
the electrode dielectric constant should then be $\epsilon _{s}\gtrsim
10-30, $ which may or may not hold for particular electrodes. This condition
has not been checked in \cite{ghosez03} and, therefore, it remains unclear
what regime corresponds to the phase transitions studied in \cite{ghosez03}.
We see that the assumption about the existence of homogeneous state made in 
\cite{ghosez03} is highly questionable and most likely the film in the
ground state would be split into domains. When the condition (\ref{eq:C1})\
is met [case (i)], the film is in a paraelectric phase when $l<l_{h}$ (see,
however, a reservation below) and becomes single domain ferroelectric in
thicker films, $l>l_{h},$ Fig. 1b. Indeed, at $l>l_{h}$ we have $\hat{A}%
-A_{d}=2L_{0}\left( \frac{1}{l_{h}}-\frac{1}{l}\right) +\frac{\lambda -L_{0}%
}{l}>0$ and the homogeneous state remains stable with regards to small
inhomogeneous fluctuations. However, it should become metastable at $l>l_{d}$%
, where domains become possible, i.e. $W\lesssim a,$ which happens at $%
l\gtrsim l_{d},$ $l_{d}\sim \sqrt{\frac{g}{\epsilon _{\perp }}}\frac{l_{h}}{%
L_{0}}\sim \frac{\lambda l_{h}}{L_{0}}>l_{h}.$ The paraelectric phase is
stable at $l<l_{h}$ with respect to small fluctuations but may, at least in
some cases, be metastable\cite{Chensky82}.

In the second case one has $L_{0}>\lambda ,$ so that $l_{d}<l_{h},$ and in
films thinner than $l_{h}$ the paraphase is already unstable with respect to
domains. Certainly, domains would form in thicker films too, so there is no
chance that the film can become homogeneously polarized. Summarizing, we
see, that the film can transform with increasing thickness (i) from
paraphase into homogeneous ferrophase at $l_{h}<l<l_{d}$ (which in some
cases may be metastable), and then into domain state at $l>l_{d}$, or (ii)
the paraphase goes over directly into domain state at $l>l_{d}.$

Now, we would like to see the effect of a finite band gap{\bf \ }$E_{g}$ and
free carriers in the ferroelectric itself on stability of homogeneous state.
One sees from Eq.~(\ref{depfield}) that at \ 
\begin{equation}
P>E_{g}/qL_{0},  \label{eq:Pguro}
\end{equation}
where $E_{g}$ is the band gap of the FE, the depolarizing field in the FE
exceeds $|E_{d}|\approx E_{g}/ql$, the band bending in the film becomes
larger than $E_{g}/2,$ and pockets with degenerate screening carriers form
in atomically thin layers at the interface\cite{guroFmet70}. We can rewrite
this into a more useful form $P>\epsilon _{s}P_{at}\kappa d_{at}(E_{g}/8\pi
E_{at}),$ where $P_{at}=q/d_{at}^{2}\sim 200\mu $C/cm$^{2}$ is the
``atomic'' polarization, $E_{at}=q^{2}/d_{at}\sim 10$eV the ``atomic''\
energy. For metallic contacts ($\epsilon _{s}\sim 1-3,$ $\kappa d_{at}\sim
1) $, and a typical band gap in FE $E_{g}=2-3$eV the condition is $P\gtrsim
10-20\mu $C/cm$^{2}$, so it seems as if it can be met in many ferroelectrics
of interest not very close to the phase transition. However, the system will
split into domains close to $T_{c}$ and is unlikely to reach such a
homogeneous state, for the same reason as in FE\ film without electrodes or
in a film with lightly doped semiconductor electrodes below.

In the FE\ slab {\em without electrodes}, the screening by the carriers is
possible only when the depolarizing field again exceeds $|E_{d}|\approx
E_{g}/ql,$ and pockets of screening{\em \ }carriers can form in atomically
thin layers at the surface of FE\cite{ivanchik61}. Before pockets form, the
depolarizing field is huge, $E_{d}=-4\pi P$ and it totally suppresses the
homogeneous polarization. Therefore, the film cannot get into monodomain
state until the absolute value of thermodynamic coercive field becomes $%
\gtrsim E_{g}/ql$. This is certainly not possible until such a low
temperature where $|A|>|A_{1}|=\left( \frac{3^{3/2}B^{1/2}E_{g}}{2ql}\right)
^{2/3}$. Using the estimate $B\sim P_{at}^{2},$ $P_{at}\sim q/d_{at}^{2},$
we obtain 
\begin{equation}
|A_{1}|\sim \left( \frac{3^{3/2}E_{g}}{2k_{B}T_{at}}\frac{d_{at}}{l}\right)
^{2/3}\sim \left( \frac{d_{at}}{l}\right) ^{2/3},  \label{eq:A1}
\end{equation}
where $k_{B}$ is the Boltzmann constant. If this state could be reached, the
system would experience a discontinuous transition into a state with large
polarization. But it could {\em not}, since domains appear much closer to $%
T_{c}.$ Indeed,\ according to Eqs.~(\ref{eq:Ad}),(\ref{eq:A1}) 
\begin{equation}
\frac{A_{d}}{A_{1}}\sim \frac{1}{\sqrt{\epsilon _{\perp }}}\left( \frac{%
d_{at}}{l}\right) ^{1/3}\ll 1.  \label{eq:dv1}
\end{equation}
The result (\ref{eq:A1})\ suggests that in typical ferroelectrics thinner
than 1-10 $\mu $m the monodomain state is simply impossible:\ the equation
of state\ has only a trivial solution $P=0$. Similar reasoning applies to
the case of {\em intrinsic semiconductor electrodes}\cite{wurfel74,batra72},
where $E_{g}$ now means the band gap of the semiconductor. These authors
have found that the transition into a homogeneous state should become first
order but did not realize that it cannot compete with domains. The above
results are in striking disagreement with claims\cite{wat} that carriers can
stabilize the homogeneous polarization in the few-nm thick FE films with
semiconductor electrodes. There is no comment on a disagreement with earlier
calculations by Ivanchik and Guro {\it et al.} \cite{ivanchik61}, who
obtained the results by solving the equation of state, accounting for
presence of the pockets with degenerate carriers that were disregarded in 
\cite{wat}.

We have shown that carriers in either ferroelectric itself or semiconductor
electrodes are usually insufficient to screen the depolarizing field not far
from the phase transition. Far from the transition other mechanisms of
screening may operate, e.g. generation of bulk/surface charged defects like
oxygen vacancies\cite{muller06def},\ surface reconstruction\cite{polarrec},
or local charge disproportionation between ion layers \cite
{muller02el,muller06def}. Yet, the formation of domains seems to be the
dominant screening mechanism in ferroelectric films, at least near the phase
transition. Our results suggest that nearly unstrained cubic ferroelectrics
go over into a multidomain state upon cooling to below the critical
temperature especially easily. In certain cases (perhaps, not observed yet)
they could transform into a monodomain state. In a multidomain state, the
electric response of the film would be determined by the properties of the
domain structure, including pinning. In this regard, it may be relevant to
the present discussion that in 12\AA\ PbTiO$_{3}$ thin films on SrTiO$_{3}$
substrate the Argonne group has detected domains \cite{Aucielo04}. In their
case the top electrode was absent, but we know that the conditions for the
domain formation in non-electroded films are similar to those in electroded
samples (capacitors) if one takes into account possible dead layers or
finite screening length in the electrodes, see above discussion of Eq.~(\ref
{eq:Ad}). Theoretically, at least, the ferroelectric domains may survive
down to thicknesses comparable to one unit cell.

APL has been partially supported by Spanish MEC (MAT2003-02600) and CAM
(S-0505/MAT/000194).

\end{document}